\pdfoutput=1
\documentclass[letterpaper,twocolumn,10pt]{article}
\usepackage{usenix}

\usepackage{tikz}
\usepackage{amsmath}

\graphicspath{{./figures/}}
\DeclareGraphicsExtensions{.pdf,.png}
\usepackage{comment}
\usepackage{lipsum}
\usepackage{color} % colors for comments
\usepackage{vwcol}
\usepackage{multirow}
\usepackage{pifont}

\usepackage{subcaption}

\usepackage{listings}
\lstdefinestyle{embedded}{
	numbers=none,
	frame=none,
	xleftmargin=0cm,
	backgroundcolor=\color{blue!3},
	framesep=1pt,
	aboveskip=3pt,
	belowskip=3pt,
	basicstyle=\small\rmfamily
}
\lstdefinestyle{small}{
	basicstyle=\linespread{0.9}\footnotesize,
}

\mathchardef\hyphenmathcode=\mathcode`\-

\let\origlstlisting=\lstlisting
\let\endoriglstlisting=\endlstlisting
\renewenvironment{lstlisting}
    {\mathcode`\-=\hyphenmathcode
     \everymath{}\mathsurround=0pt\origlstlisting}
    {\endoriglstlisting}

\usepackage{array}
\usepackage{varwidth} %for the varwidth minipage environment
\usepackage{mdwmath}
\usepackage{mdwtab}
\usepackage{float}
\newfloat{program}{t}{lop}
\usepackage{xcolor}
\usepackage{xspace}

\expandafter\def\expandafter\UrlBreaks\expandafter{\UrlBreaks\do-\do_}

\usepackage{balance}

\usepackage{makecell}
\microtypecontext{spacing=nonfrench}

\newcommand{\myparagraph}[1]{\smallskip \noindent{\bf {#1}.}}
\newcommand{\myparagraphnodot}[1]{\smallskip \noindent{\bf {#1}}}
\newcommand{\code}[1]{\mbox{\texttt{#1}}}
\newcommand{\circled}[1]{\raisebox{.5pt}{\textcircled{\raisebox{-.9pt} {#1}}}}

\newcommand{\secref}[1]{$\S$\ref{sec:#1}}
\newcommand{\subsecref}[1]{$\S$\ref{subsec:#1}}

\newcommand{\figref}[1]{Figure~\ref{fig:#1}}

\newcommand{\tabref}[1]{Table~\ref{tab:#1}}

\newcommand{\sys}{SpecFuzz}
\newcommand{\asan}{AddressSanitizer}
\newcommand{\asans}{ASan}

\newcommand{\uops}{$\mu$ops}

\usepackage{include/grumble}
\usepackage{include/report}

\begin{document}
%-------------------------------------------------------------------------------

%don't want date printed
\date{}

% make title bold and 14 pt font (Latex default is non-bold, 16 pt)
\title{\Large {\bfseries SpecFuzz} \\ {\large\normalfont Bringing Spectre-type vulnerabilities to the surface }}

% authors
\author{
{\rm Oleksii Oleksenko$^\dag$, Bohdan Trach$^\dag$, Mark Silberstein$^\ddag$, and Christof Fetzer$^\dag$}\\
{$^\dag$TU Dresden, $^\ddag$ Technion} \vspace{0.5cm}
}

\maketitle

% -*- root: main.tex -*-

\begin{abstract}

%Spectre-type attacks are a real threat to secure systems because a successful attack can undermine even an application that would be traditionally considered safe.

\sys{} is the first tool that enables dynamic testing for speculative execution vulnerabilities (e.g., Spectre).
The key is a novel concept of \emph{speculation exposure}: The program is instrumented to simulate speculative execution in software by forcefully executing the code paths that could be triggered due to mispredictions, thereby making the speculative memory accesses visible to integrity checkers (e.g., AddressSanitizer).
Combined with the conventional fuzzing techniques, speculation exposure enables more precise identification of potential vulnerabilities compared to state-of-the-art static analyzers.

Our prototype for detecting Spectre V1 vulnerabilities successfully identifies all known variations of Spectre V1 and decreases the mitigation overheads %compared to the deployed Speculative Load Hardening mitigation
 across the evaluated applications, reducing the amount of instrumented branches by up to 77\% given a sufficient test coverage.

\end{abstract}

% -*- root: main.tex -*-

\section{Introduction}
\label{sec:introduction}

Spectre~\cite{Kocher18, Koruyeh2018, Google2018, schwarz2018netspectre} is a
class of 
%microarchitectural 
attacks that poses a significant threat to system security.
It is a \emph{microarchitectural attack}, an attack where a malicious actor extracts
secrets by exploiting security flaws in the CPU architecture rather than in
software. %instead of the software that runs on it.
Such attacks are particularly dangerous as they compromise the security of bug-free programs. % that would normally be considered bug-free.

\emph{Spectre-type} microarchitectural attacks exploit branch speculations to
access victim's memory. 
%of a branch to leak 
%driven by hardware optimizations. %that drive speculative execution.
For example, if an array access is guarded by an index bounds check, the CPU branch predictor 
might speculate that the check will pass and thus perform the memory access
before the index is validated. If the speculation turns out to be wrong, 
the CPU rolls back the respective changes in the architectural state (e.g., in registers),
but it does not cleanse its microarchitectural state (e.g., cached data).
Spectre-type attacks use this property to exfiltrate the results of computations executed on this \emph{mispredicted path}.
% \al{Sorry, I like it better this way}
%For example, the
%speculative execution may alter data residency in hardware caches, and some Spectre-type attacks
%use this property to exfiltrate the data computed on the \emph{mispredicted
%path}.

Unfortunately, many variants of Spectre hardware vulnerabilities are not expected to be fixed by hardware vendors, most notably Intel~\cite{Intel2019}.
%Many speculative execution attacks, e.g., Meltdown~\cite{Lipp18},
%Foreshadow~\cite{Bulck2018}, MDS~\cite{RIDL,Fallout,ZombieLoad}, are expected to
%be fixed by hardware vendors. 
%
Therefore, the burden of protecting programs lies entirely on software developers~\cite{Mcilroy19}.

This observation led to the development of software tools for Spectre mitigation. 
They identify the code snippets purported to be vulnerable to the Spectre
attacks and instrument them to prevent or eliminate unsafe speculation. 
Inherently, the instrumentation incurs runtime overheads, thereby leading to the
apparent tradeoff between security and performance. 

Currently, all the existing tools exercise only the \emph{extreme} points in
this tradeoff, offering either poor performance with high security, or poor security
with high performance.

Specifically, conservative techniques~\cite{SLH,Intel18a,Retpoline, SigArchPost}
pessimistically harden every \emph{speculatable instruction} (e.g., every
conditional branch) to either prevent the speculation or make it provably
benign. This approach is secure, but may significantly hurt program performance~\cite{YSNB}.

On the other hand, static analysis tools~\cite{MSVC, RHScanner,
Respectre} reduce the performance costs by instrumenting only known \emph{Spectre gadgets}---the code
patterns that are typical for the attacks. However, the analysis is
imprecise and may overlook vulnerabilities, either because the vulnerable code does not match the
expected patterns~\cite{Kocher2018a}, or due to the limitations of the analysis
itself (e.g., considers each function only in isolation).

We seek to build a tool that exercises a different point on the
security-performance tradeoff curve by eliding unnecessary
%instrumentation without compromising security. % \al{Too strong of a claim. SpecFuzz does compromise security}
instrumentation without restricting ourselves to specific gadgets.
Arguably, a key challenge is to precisely identify vulnerable code regions, yet
this task is hard to achieve via static analysis. Instead, in this work we harness \emph{dynamic} testing (e.g., fuzzing) 
to detect Spectre-type vulnerabilities. 

Fuzzing~\cite{fuzzingbook2019} is a well-established testing technique.
The basic idea of fuzzing is simple: Add integrity checks to the tested software (e.g., with AddressSanitizer~\cite{Serebryany2012}) and feed it with randomized inputs to find cases that trigger a bug.
This technique is commonly used to detect stability issues and memory errors~\cite{OSSFuzz}.

In principle, Spectre-type attacks effectively perform unauthorized accesses to data via
out-of-bound reads, thus they are supposed to be caught via fuzzing. 
Unfortunately, this is not the case because the accesses are invoked
\emph{speculatively, on a mispredicted path}, therefore are discarded by
hardware without being exposed to software. As a result, they remain invisible to
runtime integrity checkers.

%To overcome this problem, we rely on simulation. 
We introduce \emph{speculation exposure}, the first technique to enable dynamic testing for Spectre-type vulnerabilities.
Speculation exposure leverages \emph{software} simulation of speculative paths to turn speculative vulnerabilities into conventional ones and,
thus, make them detectable by memory safety checkers.
The concept is generic and can be applied to different Spectre attacks.% with only small modifications.

Speculation exposure consists of four phases executed for every speculatable instruction:
\circled{1} take a checkpoint of the process state,
\circled{2} simulate a misprediction,
\circled{3} execute the speculative path, and
\circled{4} rollback the process to the checkpoint and continue normal execution.
This way, we temporarily redirect the normal application flow into the
speculative path so that all invalid memory accesses on it become
visible to software. This method simulates the worst-case scenario by
examining each possible mispredicted path, without making assumptions about the
way the underlying hardware decides whether to speculate or not.

We further extend speculation exposure to \emph{nested speculation}, which
occurs when a CPU begins a new speculation before resolving the previous one.
To simulate it, for each speculatable instruction, we dynamically generate a \emph{tree}
of all possible speculative paths starting from this instruction and branching on every next speculatable instruction.
The complete nested simulation, however, has proven to be too slow.
To make fuzzing practical we develop a heuristic which prioritizes traversal of
the speculation sub-trees with high likelihood of detecting new vulnerabilities.

To showcase our method, we implement \sys{}, a tool for detecting Bounds Check Bypass (BCB) vulnerabilities.
\sys{} simulates conditional jump mispredictions by placing an additional jump
with an inverted condition before every conditional jump. During the simulation 
it executes the inverted jump and then rolls back to return to the original control flow.
To detect invalid accesses on the simulated speculative path, \sys{} relies on \asan{}~\cite{Serebryany2012}.

\sys{} may serve as a tool for both offensive and defensive security.
For the former (e.g., penetration testing), it finds vulnerabilities in software, records their parameters, and generates test cases.
For the latter, the fuzzing results are passed to automated hardening tools
(e.g., Speculative Load Hardening~\cite{SLH}) to elide unnecessary
instrumentation of the instructions deemed safe. Note that
the code not covered by fuzzing remains instrumented and protected
conservatively as before, hence lower fuzzing coverage might affect performance but not security.

Our evaluation shows that \sys{} successfully detects vulnerable gadgets in all test programs.
It detects more potential vulnerabilities than the state-of-the-art and reduces
the overheads of conservative instrumentation of all conditional branches.
For example, it elides the instrumentation from about a half of branches in the security-focused libHTP library, and improves the performance of hardened OpenSSL RSA function, resulting in only 3\% slowdown over its vanilla version, compared to the 22\% slower conservative hardening.

\vspace{0.2cm}

Our contributions include:
\begin{itemize}
    \setlength{\parskip}{0pt}
    \setlength{\itemsep}{4pt plus 1pt}
    \item Speculation exposure, a generic simulation method for Spectre-type vulnerabilities that makes them detectable through dynamic testing.
    \item \sys{}, an implementation of the method applied to detection of Bounds Check Bypass vulnerabilities.
    \item A fuzzing strategy that makes nested speculative exposure feasible by prioritizing the paths that are the most likely to contain vulnerabilities.
    \item An analysis technique for processing and ranking the results of dynamic testing with \sys{}.
    \item Evaluation of \sys{} on a set of popular libraries.
\end{itemize}

% -*- root: main.tex -*-

% -*- root: ../main.tex -*-

\begin{figure}[t]
    \begin{lstlisting}[frame=tb]
i = input[0];
if (i < size) {
    secret = foo[i];
    baz = bar[secret]; }\end{lstlisting}
    \caption{A potential Bounds Check Bypass vulnerability.}
    \label{fig:bcb_pattern}
\end{figure}

\section{Background}
\label{sec:background}

\subsection{Speculative Execution and Attacks}
\label{subsec:spec-execution}

\myparagraph{Speculative Execution}
In modern processors, execution of a single instruction is carried out in several stages, such as fetching, decoding, and reading.
To improve performance, nearly all modern CPUs execute them in a pipelined fashion:
When one instruction passes a stage, the next instruction can enter the stage without waiting for the first one to pass all the following stages.
This allows for much higher levels of instruction parallelism and for better utilization of the hardware resources.

However, in certain situations---called hazards---it is not possible to begin executing the next instruction immediately.
A hazard may happen in three cases: a structural hazard appears when there are no available execution units,
a data hazard---when there is a data dependency between the instructions,
and control hazard---when the first instruction modifies the control flow (e.g., at a conditional branch) and the CPU does not know what instruction will run next.
As the hazards are stalling the CPU, they can significantly reduce its performance.

To deal with control hazards (and sometimes, with data hazards), modern CPUs try to predict the outcome of the situation and start \textit{speculatively executing} the instructions assumed next.
For example, when the CPU encounters an indirect jump, it predicts the jump target based on the history of recently used targets and redirects the control flow to it.
While the CPU does not know if the prediction was correct, it keeps track of the speculative instructions in a temporary storage, called \emph{Reorder Buffer} (ROB).
The results of these speculative computations are kept in internal buffers or registers and are architecturally invisible (i.e., the software does not have access to them).
Eventually, the CPU resolves the hazard and, depending on the outcome, either commits the results to the architectural state or discards them.

\myparagraph{Speculative Execution Attacks}
%\label{subsec:transient-attacks}
In a speculative execution attack (in short, \emph{speculative attack}), the attacker intentionally forces the CPU into making a wrong prediction and executing a wrong speculative path (i.e., executing a \emph{mispredicted path}).
Because taking the path violates the application semantics, it may bypass security checks within the application.
Moreover, should any exceptions appear on the mispredicted path, they will be handled only during the last pipeline stage (retirement).

For a long time, this behavior was considered safe because the CPU never commits the results of a wrong speculation.
However, as the authors of Spectre~\cite{Kocher18} and Meltdown~\cite{Lipp18} discovered, some traces of speculative execution are visible on the microarchitectural level.
For example, the data loaded on the mispredicted path will not show up in the CPU registers, but will be cached in the CPU caches.
The attacker can later launch a side-channel attack~\cite{Tromer10,Yarom14} to retrieve the traces and, based on them, deduce the speculative results.

\myparagraph{Bounds Check Bypass}
%\label{subsec:bcb}
In this paper, we will showcase our dynamic testing technique on one of the speculative attacks---Bounds Check Bypass (BCB, also called Spectre v1)~\cite{Kocher18}.
In essence, BCB is a conventional out-of-bounds memory access (e.g., buffer overflow) that happens on a mispredicted path, triggered by a wrong prediction of a conditional jump.

Consider the code snippet in \figref{bcb_pattern}.
Assuming that the attacker can control the \code{input} value, she can send several in-bounds inputs that would train the branch predictor to anticipate that the check at line~2 will pass.
Then, the attacker sends an out-of-bounds input, the branch predictor makes a wrong prediction, and the program speculatively executes lines~3--4 even though the program's semantics forbid so.
It causes a speculative buffer overread at line~3 and the read value is used as an index at line~4.

Later, the CPU finds out that the prediction was wrong and discards the speculated load, but not its cache traces.
The adversary can access the traces by launching a side-channel attack and use them to deduce the \code{secret} value:
The address read at line~4 depends on the \code{secret} and, correspondingly, finding out which cache line was used for this memory access allows the attacker to also find out the \code{secret} value loaded on the speculative path.

Note that without the bounds check at line~2, this vulnerability would be a conventional buffer overflow which can be detected by memory safety techniques, such as AddressSanitizer~\cite{Serebryany2012} or Intel MPX~\cite{intelsys}.
However, since the CPU cancels the speculation after detecting a misprediction, these techniques turn ineffective.

\subsection{Fuzzing}
\label{subsec:fuzzing}

Fuzzing is a technique for discovering bugs and vulnerabilities in software by exposing it to diverse conditions and inputs.
A fuzzing tool (\emph{fuzzer}) automatically generates randomized inputs either from scratch, based on input grammars, or by mutating an existing \emph{input corpus}.
The fuzzer then feeds these inputs to the application and monitors its behavior: If an abnormal behavior (e.g., a crash) is observed, the fuzzer reports a bug.
Since many bugs do not manifest themselves in externally-visible failures, fuzzing is often used in combination with memory safety techniques that can detect internal errors.

One important parameter of fuzzing is its \emph{coverage}, which indicates how extensively the software was tested during fuzzing.
Coverage can be defined in many ways, but the most common is to define it as a
ratio of the control-flow graph edges that were executed at least once during fuzzing to the total number of edges in the application.
Coverage mainly depends on the effectiveness of the input generator, that is, on how effectively it can generate inputs that trigger new control-flow paths.
It is also highly dependent on the quality of the \emph{fuzzing driver}, the wrapper that interfaces the application to the fuzzer.
If the driver does not call some of the application's functions, they will never be covered by fuzzing, regardless of how effective the generator is.

% -*- root: main.tex -*-

\begin{figure}[t]
    \centering
    \includegraphics[scale=0.4]{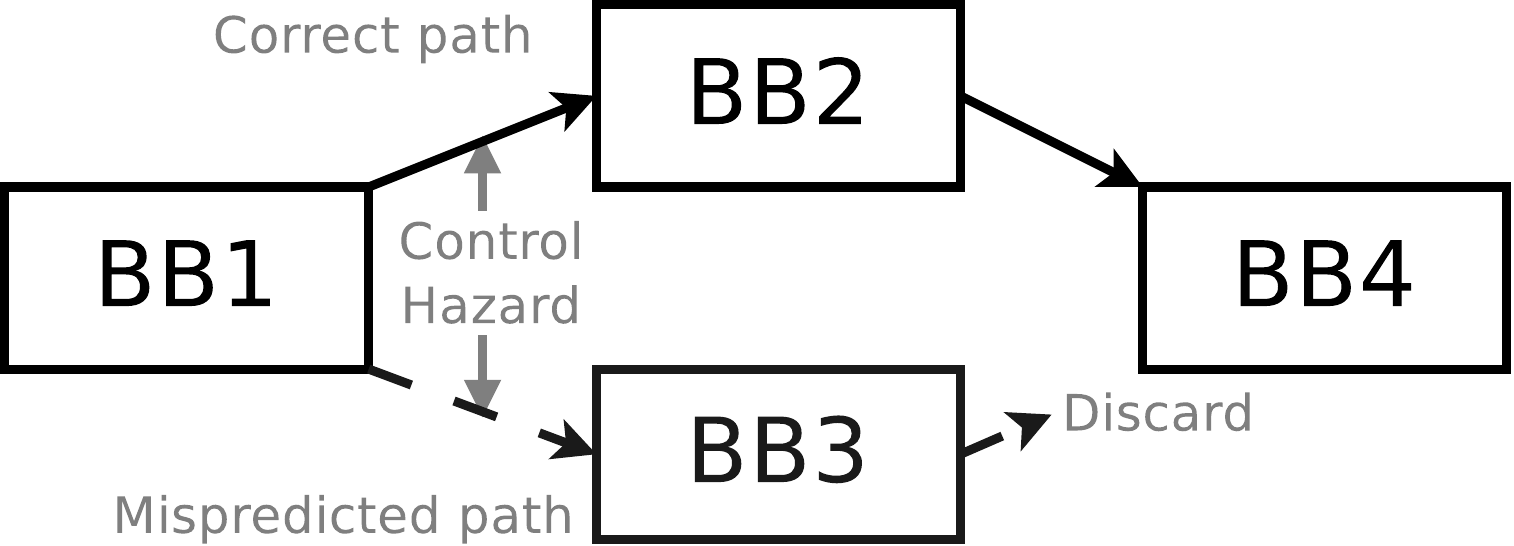}
    \caption{Speculative execution.
    Due to a misprediction, the program executes basic blocks BB3 and BB4, then detects the mistake, discards the results, and continues execution starting from BB2.}
    \label{fig:misprediction-example}
\end{figure}

\section{Speculation Exposure}
\label{sec:bigidea}

\al{Note to myself: This paragraph basically repeats intro. Shorten it}
Speculative vulnerabilities are notoriously hard to find because hardware strives to hide the effects of speculative execution from software, making it impossible to detect such vulnerabilities with conventional testing methods.
In this paper, we approach the problem by simulating the unsafe hardware optimization in software.
We call this approach \emph{speculation exposure}.

To understand how we construct the simulation, first consider how speculative execution is implemented in hardware~(\subsecref{spec-execution}).
When a hazard appears (e.g., at a conditional or an indirect jump), the CPU
\circled{1} makes a prediction of its outcome,
\circled{2} executes the speculative path while temporarily keeping the results in internal buffers,
\circled{3} eventually eliminates the hazard and either commits the results (correct prediction) or discards them (wrong prediction), and
\circled{4} proceeds with the correct path.

For example, in \figref{misprediction-example}, the CPU might make a wrong prediction that BB1 (Basic Block 1) will proceed into BB3.
It will start executing BB3, BB4, and maybe even further, depending on how long it takes to resolve the hazard.
When the hazard is resolved, the CPU determines that the prediction was wrong and discards all changes made on the speculative path.
Afterward, it redirects the control flow to the correct path and proceeds with the execution starting from BB2.

The core idea behind speculation exposure is to simulate this behavior in software with a \emph{checkpoint-mispredict-rollback} scheme:
At a potential hazard, we \circled{1} take a checkpoint of the current process state.
%\ms{process or processor?} \al{process (OS abstraction, not CPU) - registers and memory contents}
Then, we \circled{2} diverge the control flow into a wrong (mispredicted) path and start executing it.
When a termination condition is reached (e.g., a serializing instruction is executed), we \circled{3} rollback to the checkpoint and \circled{4} proceed with normal execution.
The pattern can be applied to data hazards too:
Instead of diverging the control flow, we would replace a memory/register value with a mispredicted one.

This basic mechanism simulates the worst case scenario when a CPU always mispredicts and always speculates to the greatest possible depth.
Such a pessimistic approach makes the testing results universally applicable to different CPU models and any execution conditions.
Moreover, it also covers all possible combinations of correct and incorrect predictions that could happen at runtime (see~\subsecref{idea-nesting}).

\subsection{Components of Speculation Exposure}
\label{subsec:components}

There are four core components: a checkpointing mechanism, a simulation of
mispredictions, a detection of faults on the simulated path, and a mechanism for detecting termination conditions.

\myparagraph{Checkpointing}
For storing the process state, we could use any of the existing checkpointing
mechanisms, ranging from full-process checkpoint (e.g., CRIU~\cite{CRIU}) to
transactional memory techniques (e.g., Intel TSX~\cite{intelsys}). However,
checkpointing is on the critical path in our case, thus heavy-weight
mechanism would either increase the testing time, or reduce the number of
inputs used in fuzzing under a fixed time budget.
We describe the checkpointing mechanism used in our implementation in \subsecref{basic-sim}.

\myparagraph{Simulating Misprediction}
To simulate misprediction, we instrument basic blocks in a way that forces control flow to enter the paths that the CPU would otherwise take speculatively.
The nature of the instrumentation depends on the exact type of the speculative
execution attack being simulated (see \secref{specfuzz} and \secref{apply-simul-spectre} for a
detailed discussion about applying this technique to different Spectre attacks).

\myparagraph{Detection of Vulnerabilities}
In Spectre-type attacks, the data is leaked when a program speculatively reads from or writes to a wrong object.
Therefore, when we have a mechanism for simulating speculative execution, the detection of actual vulnerabilities boils down to the conventional memory safety problem;
detecting bounds violations.
This is a well-developed field with many existing solutions \cite{softbound09,intelsys,Serebryany2012}.
In this work, we rely on AddressSanitizer~\cite{Serebryany2012}.

\myparagraph{Terminating Simulation}
The simulation mimics the termination of the speculative execution by hardware.
Speculative execution terminates: \emph{(i)} upon certain
\emph{serializing}  instructions (e.g., \code{LFENCE}, \code{CPUID},
\code{SYSCALL}, as listed in the CPU documentation~\cite{intelsys}), and \emph{(ii)} after the speculation exhausts certain
hardware resources. Thus, the simulation terminates when one of those
conditions is satisfied.

Note that terminating the simulation earlier results in faster fuzzing and could be used as an optimization, but it could miss vulnerabilities.
Below we discuss the hardware resources used in speculation to determine the simulation termination conditions.

\subsubsection{Termination conditions}

All program state changes made during the speculative execution must be temporarily stored in internal hardware buffers,
so that they can be reverted if the prediction is incorrect.
Accordingly, once at least one of these buffers becomes full the speculation
stops.

On modern Intel CPUs, there are several buffers that can be exhausted~\cite{intelsys}: Reorder
Buffer (ROB), Branch Order Buffer (BOB), Load Buffer (LB), Store Buffer (SB),
Reservation Station (RS), Load Matrix (LM), and Physical Register Reclaim Table
(PRRT). We seek to find the one that overflows first.

LM and PRRT are not documented by Intel. LB, SB, and RS are also not useful for practical simulations as their entries
could be reclaimed dynamically (policy is undocumented) during speculative
execution. Therefore, we do not simulate these buffers and assume that they do
not restrict the depth of the speculation.
%Correctly simulating them would require an often-undocumented information about the reclamation algorithms in Intel CPUs.

We are left with ROB, which keeps track of all speculative microoperations (\uops), and BOB, which tracks unresolved branch predictions.
We choose ROB because BOB is not portable
as it is a specific optimization of Intel CPUs~\cite{bob6799268}.
%\ms{can you substantiate by a reference?}
%\al{It is a recent, patented technology. I know that it is available only on Intel CPUs, but finding a hard proof for it is tricky. I'll skip now and find something for the iteration}

In Intel x86, any speculative path can contain at most as many \uops{} as there are
entries in ROB\footnotemark.
In modern CPUs, its size is under 250 \uops{} (the largest we know is 224 entries, on Intel Skylake architecture~\cite{intelOptRef}).

\footnotetext{Some CPU architectures (e.g., CPR~\cite{Akkary03}) could speculate beyond the ROB size.
However, to the best of our knowledge, that is not the case for the existing x86 CPUs}

The simulation terminates after reaching 250 instructions, which is a
conservative estimate because one instruction is typically mapped into one or more \uops{}.
The only exception is \uops{} fusion, when CPU merges several instructions into one.
However, on Intel CPUs, it is limited to a small set of instruction combinations~\cite{intelOptRef}.
To account for this effect, we count these combinations as a single instruction.

Note that a tighter bound on the number of speculated instructions (e.g., through simulation of a smaller buffer) could have
improved the fuzzing time without affecting correctness.

%Note, that ithe choice of the buffer does not impact the correctness of the simulation:
%If one of the buffers can be filled sooner than ROB, then our simulation would simply overapproximate the behaviour of the real CPU\@.
%In the worst case, it would lead to false positives.

\begin{figure}[t]
    \centering
    \includegraphics[scale=0.47]{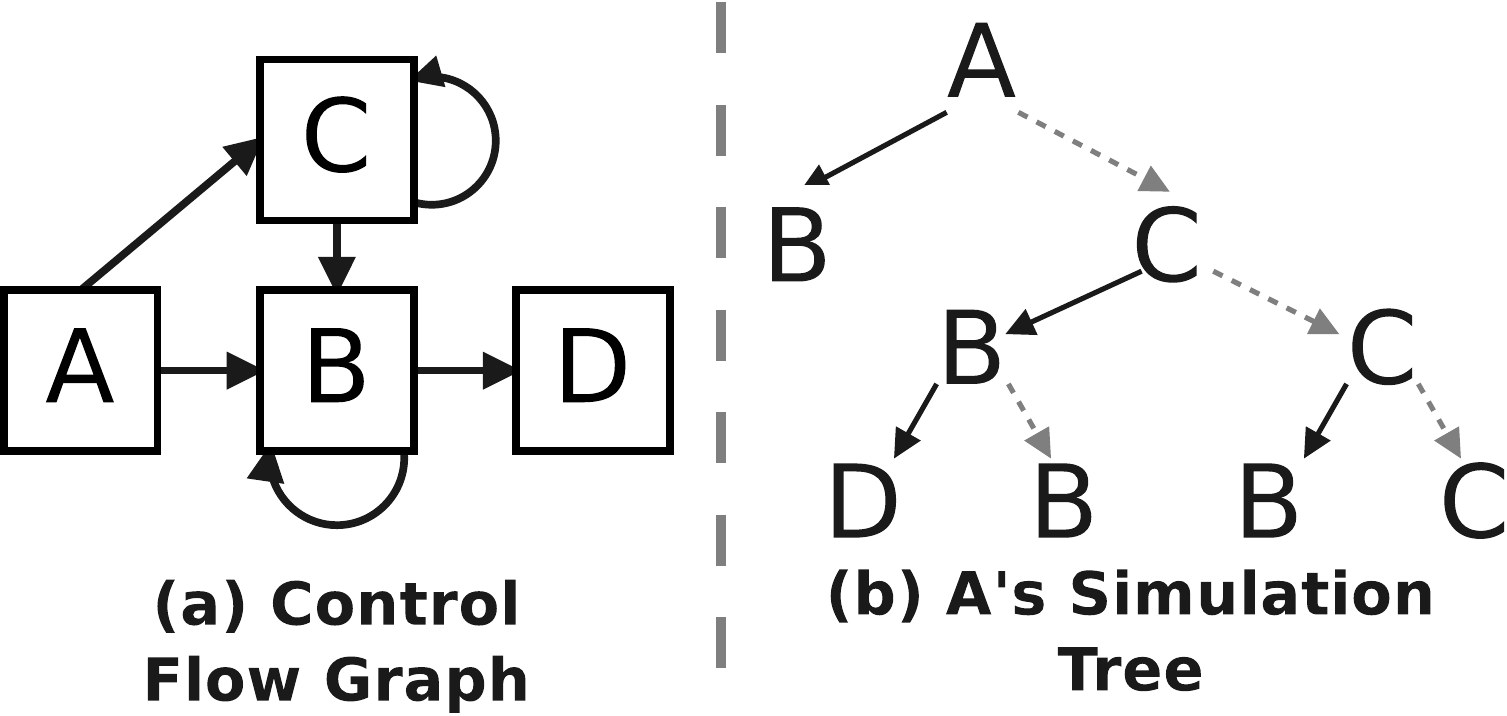}
    \caption{Nested speculation exposure for the flow A$\rightarrow$B$\rightarrow$D.
    Dashed lines are mispredicted speculative paths.}
    \label{fig:nesting}
\end{figure}

\subsection{Nested Speculation Exposure}
\label{subsec:idea-nesting}

The CPU may perform \emph{nested} speculation; that is, it can make a prediction while
already executing a speculative path. Since we do not make any assumptions about the predictions,
every speculatable instruction triggers not a single simulation, but a series of \emph{nested simulations}.
We refer to a tree of all possible speculative paths as a \emph{simulation
tree}. A simulation tree for each speculatable instruction is regenerated for each program input.

Instead of traversing the complete simulation tree (\emph{complete simulation}), we could simulate only a subset of all mispredictions.
Then, an \emph{order of a simulation} is the maximum number of nested mispredictions it simulates.
In other words, an order is the maximum depth of the simulation tree.
Accordingly, an \emph{order of a vulnerability} is defined as the minimum order
of a simulation that triggers this vulnerability. An \emph{order of a
speculative path} is the number of mispredictions required to enter it.

Consider the example in \figref{nesting}.
The left side (\figref{nesting}a) is a control-flow graph.
%(we start a new simulation before every branch) 
Suppose that the correct flow is \code{ABD}.

If we simulate branch mispredictions, then the simulation tree of branch \code{A} would be as shown in \figref{nesting}b.
The simulation of order 1 for that branch traverses only the path \code{(ACBD)},
simulating only the first misprediction, and then following the original flow
graph.  
The simulation of order 3 would traverse three additional paths: \code{ACBB},
\code{ACCB} and \code{ACCC}, according to misspeculation of A and B; A and C;
and A,C and C respectively. The four paths constitute a complete simulation tree of the branch \code{A}.
Every branch (or, more generally, every speculatable instruction) has its own simulation tree and the tree 
has to be traversed every time the branch is executed.

%\ms we said that earlier
%Conceptually, the only difference between the basic (order 1) and the nested
%versions of speculation exposure is that the nested version can simulate a new even if we are already on a speculative path.
%Otherwise, the checkpointing and the simulation mechanisms, as well as the termination conditions stay the same.
%However, 
Nested simulation dramatically increases the fuzzing time. However, in \sys{} we
use a heuristic which, while traversing only a small portion of the speculation
tree on each input, shows high detection rates.
% only slightly lower than the exhaustive search. \al{not lower at all. With prioritized simulation, we found everything that complete simulation had found, and even some more}
We discuss it in detail in \subsecref{nesting}.

% -*- root: main.tex -*-
\section{SpecFuzz}
\label{sec:specfuzz}

To showcase speculative exposure on a specific class of vulnerabilities, we develop \sys{}, a tool for simulating and detecting Bounds Check Bypass (BCB)~\cite{Kocher18}.
We discuss other Spectre-type attacks in~\secref{apply-simul-spectre}.

As described in \subsecref{spec-execution}, BCB in its core contains a speculative out-of-bounds access caused by a conditional jump misprediction.
To expose such accesses, we create a modified (instrumented) version of the application which executes not only the normal control flow but also enters all possible speculative paths.

\sys{} works as follows (see \figref{instrumentation_example}):
Before every conditional branch (line 4), it inserts a call to a checkpointing function (line 1) that stores the process state and initializes simulation.
Then, it adds a sequence of instructions that simulate a misprediction (lines 2--3) and force the control flow into the mispredicted path.
Specifically, \sys{} inserts a jump with an inverted condition (line 2), followed by a jump into the body of the conditional block, thus skipping the original branch (line 3).
During the simulation, \sys{} periodically checks if a termination condition has appeared (line 8).
If the check passes, \sys{} restores the process state from the previous checkpoint (line 9) and continues the program execution.

We implement this design as a combination of an LLVM~\cite{LLVM2004} compiler backend pass for the x86 architecture and a runtime library. % (in total, over 2000 LoC).

\subsection{Basic Simulation}
\label{subsec:basic-sim}

% -*- root: ../main.tex -*-

\begin{figure}[t]
    \begin{minipage}[t]{0.4\columnwidth}
    \begin{lstlisting}[frame=tb]



if x < array_size:

    result = array[x]
    ...


\end{lstlisting}
    \subcaption{Native version}
    \end{minipage}
    \begin{minipage}[t]{0.58\columnwidth}
    \begin{lstlisting}[frame=t,backgroundcolor=\color{blue!10},numbers=none]
    checkpoint()
    if x >= array_size:
        goto skip_branch
\end{lstlisting}\begin{lstlisting}[frame=none,numbers=none]
    if x < array_size
\end{lstlisting}\begin{lstlisting}[frame=none,backgroundcolor=\color{blue!10},numbers=none]
skip_branch:
\end{lstlisting}\begin{lstlisting}[frame=none,numbers=none]
    result = array[x]
    ...
\end{lstlisting}\begin{lstlisting}[frame=b,backgroundcolor=\color{blue!10},numbers=none]
    if terminate_simulation():
        rollback() // to line 4
\end{lstlisting}
    \subcaption{Simulation of conditional branch misprediction}
    \end{minipage}

    \caption{\sys{} instrumentation.}
    \label{fig:instrumentation_example}
\end{figure}

\begin{figure}[t]
    \centering
    \includegraphics[scale=0.18]{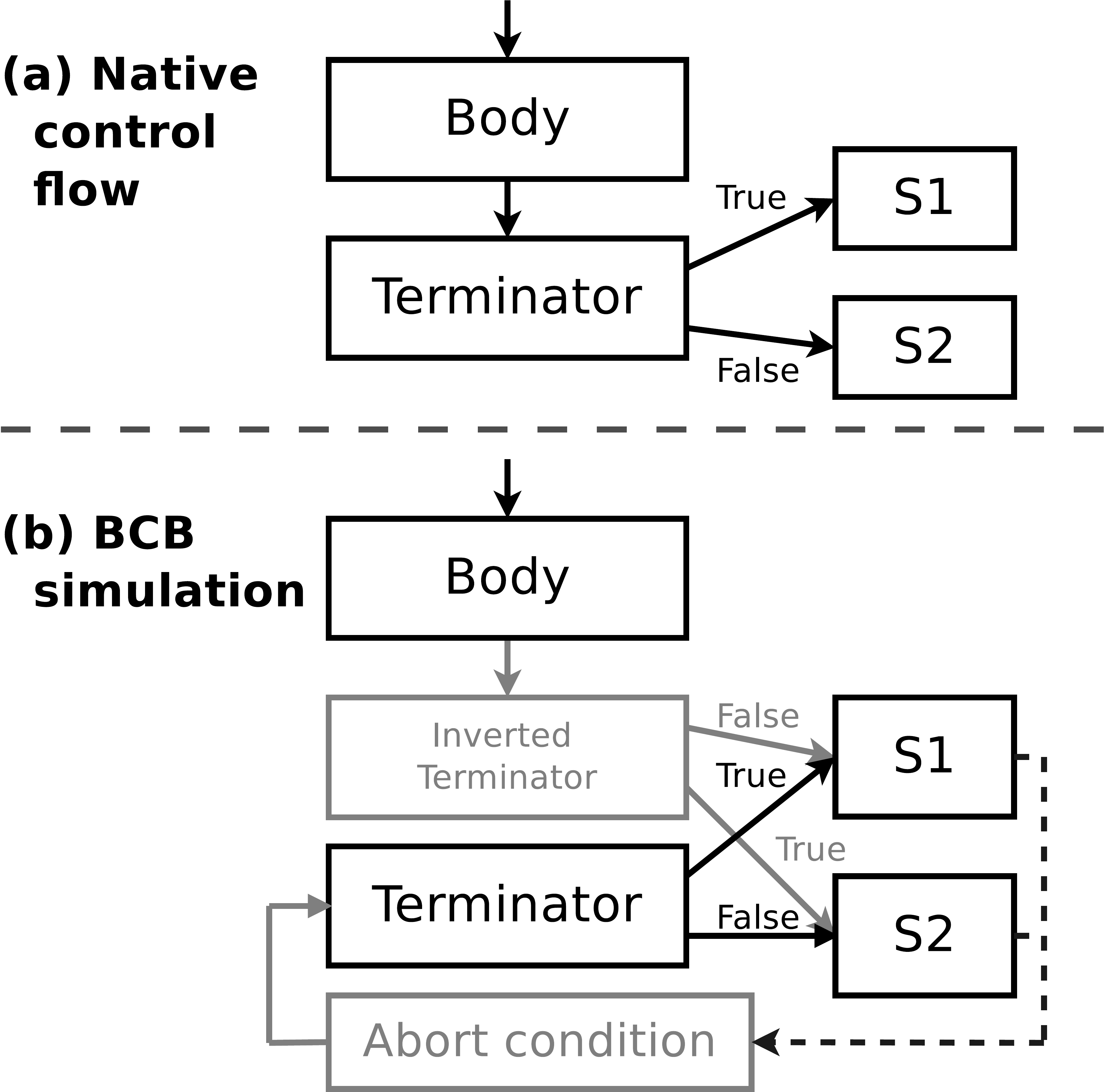}
    \caption{Simulation of conditional branch mispredictions:
    On simulated speculative paths, all conditional terminators are replaced by terminators with inverse conditions.}
    \label{fig:bcb-simulation}
\end{figure}

\myparagraph{Simulating Branch Misprediction}
%\label{subsec:misprediction}
~\sys{} simulates mispredictions by forcing the application into taking a wrong branch at every conditional jump.
We implement this behavior by replacing all conditional terminators in the program with the ones that have an inverted condition (see \figref{bcb-simulation}).
Now, when the original basic block (BB) would proceed into the successor $S1$, the modified terminator diverges the control flow into $S2$.
The original terminator is moved into a separate BB, and the control flow returns to normal execution by rolling back into this BB after the simulation.

As a result, every time the program reaches this BB, it first executes the simulated path, then rolls back to the BB and continues with normal execution.
%We apply this instrumentation to all conditional branches.

\myparagraph{Saving and Restoring Process State}
%\label{subsec:checkpointing-mechanism}
The main requirement to the rollback mechanism used in \sys{} was to have low performance impact so that the fuzzing time is kept short.
To this end, we implement a light-weight in-process mechanism that snapshots the CPU state before starting a simulation and records the memory changes during the simulation.

To store the CPU state, we add a call to a checkpointing function (a part of the runtime library) before every conditional jump.
The function takes a snapshot of the register values (including GPRs, flags, SIMD, floating-point registers, etc.) and stores it into memory.
During the rollback, we restore the register values based on the snapshot.
The function also stores the address of the original conditional jump (i.e., original terminator) that we later use as a rollback address.

This approach, however, is not efficient when applied to saving the memory state because it would require dumping the memory contents into disk at every conditional jump.
To avoid the performance overhead linked with this expensive operation, we instead rely on logging the memory changes that happen during the simulation.
Before every instruction that modifies memory (e.g., \code{mov}, \code{push}, \code{call}), we store the address it modifies and its previous value onto a stack-like data structure.
Then, to do a rollback, we go through this data structure in the reverse order and restore the previous memory values.

Currently, \sys{} supports only fixed-width writes;
If the pass encounters \code{REP MOV}, compilation fails with an error.
Yet, we did not encounter any issues with that during our experiments because Clang in its default configuration does not use these instructions.

\myparagraph{Detecting and Handling Errors}
%\label{subsec:handling-errors}
With the simulation mechanism at hand, we now need a mechanism to detect invalid accesses on speculative paths.
In \sys{}, we utilize \asan{}~\cite{Serebryany2012} (\asans{}) to detect out-of-bounds accesses and a custom signal handler to handle the errors that inevitably appear during the simulations.

We had to modify the behavior of \asans{} to our needs.
In contrast to normal, non-speculative execution, the process does not crash if an error happens during the speculation.
Instead, the CPU silences the error by discarding its effects when the misprediction is detected.
To simulate this behavior in \sys{}, we adjusted the error response mechanism in \asans{} to record the violation in a log and continue the simulation.
Accordingly, one test run might detect several (sometimes, hundreds of) violations.

Similarly, we have to recover from runtime faults.
We register a custom signal handler that logs and rolls back after the signals that could be caused by an out-of-bounds access, such as \code{SIGSEGV} and \code{SIGBUS}.
We also rollback after other faults (e.g., division by zero), but we do not record them in the log as they are irrelevant to the BCB vulnerability.
We perform an immediate rollback because hardware exceptions are supposed to terminate speculative execution.
Even though on some CPU models exceptions may not terminate speculation (see
Meltdown-type attacks~\cite{Lipp18,Canella2018}), we ignore such cases assuming
they will be fixed at the hardware level similarly to Meltdown.
%consider it as a CPU implementation bug that should be fixed on the hardware level.

\myparagraph{Terminating Simulation}
%\label{subsec:terminating-simulation}
As discussed in \secref{bigidea}, we terminate the simulation either when we encounter a serializing instruction or when the maximum depth of speculation is reached.

To implement the first case, we simply insert a call to the rollback function before every serializing instruction.
As serializing, we consider the instructions listed as such in the Intel documentation~\cite{intelsys} (e.g., \code{LFENCE}, \code{CPUID}, \code{SYSCALL}).

To count instructions at runtime, we keep a global instruction counter and set it to zero when a simulation begins.
At the beginning of every basic block, we add its length to the counter.
(We know the length at compile time because \sys{} is a backend pass).
When the counter value reaches 250 (maximum ROB size, see \secref{bigidea}), we invoke the rollback function.

% -*- root: ../main.tex -*-
\begin{table}[]
    \center
    \setlength{\tabcolsep}{3.3pt}
    \renewcommand{\arraystretch}{1}
    \begin{tabular}{|l|cccccc|}
        \hline
        Order & JSMN & Brotli & HTTP & libHTP & YAML & SSL \\
        \hline
        1 & 6 & 74 & 6 & 221 & 77 & 1254 \\
        2 & 5 &  9 & 4 &  64 & 92 &  366 \\
        3 & 7 & 12 & 2 &  33 & 14 &  253 \\
        4 & 1 &  6 & 3 &   5 & 16 &   91 \\
        5 & 1 &  2 & 1 &   2 &  6 &   - \\
        6 & 0 &  0 & 0 &   2 &  2 &   - \\
        \hline
        Total & 20 & 103 & 16 & 327 & 207 & 1964 \\
        \hline
        Iterations & 933 & 3252 & 1582 & 540 & 1040 & 227 \\
        \hline
    \end{tabular}
%     \vspace{-1mm}
    \caption{Distribution (by order) of the vulnerabilities detected by 24 hours of fuzzing non-prioritized 6th-order simulation.
    This experiment motivates prioritized simulation: Even though all fuzzing rounds simulated all 6 orders of misprediction, most of the detected vulnerabilities required only a few mispredictions.
    Since execution of OpenSSL was too slow, we simulated it only to the 4th order. }
    \label{tab:nesting}
    % \vspace{-4mm}
\end{table}

\subsection{Nested Simulation}
\label{subsec:nesting}

To implement nested simulation, we maintain a stack of checkpoints:
Every time we encounter a conditional branch, we push the checkpoint on the stack, as well as the current value of the instruction counter and a pointer to the previous stack frame.
All later writes will be logged into the new stack frame.
At rollback, we restore the topmost checkpoint and revoke the corresponding memory changes.
This way, \sys{} traverses all possible combinations of correct and incorrect predictions in the depth-first fashion.

%\todo{The text partially duplicates \secref{bigidea}. Maybe, remove or rewrite}
%Let us come back to the example of a simulation tree in \figref{nesting}b.
%Here, \sys{} would push the first checkpoint before the branch \code{A}, diverge the control flow into \code{C}, and record all the memory changes happening in \code{C} into this stack frame.
%Then, at branch \code{C}, it would push another checkpoint on the stack and proceed executing \code{C} the second time, and then repeat the procedure the third time until reaching the threshold of 250 instructions (to remind, we assume that all basic blocks in this example are 100 instructions long).
%Now, when we have reached a termination condition, we pop the topmost checkpoint of the stack and roll back to the second execution of \code{C} and continue into \code{B}.
%After \code{B}, we roll back to the second execution of \code{C} and continue into \code{B}.
%We proceed with this algorithm until traversing the whole simulation tree.

\myparagraph{Coverage Trade-off}
The number of paths to traverse increases exponentially with the order of the
simulation. In most programs, the density of conditional branches is approximately one in ten instructions.
If we assume the maximum depth of speculative execution to be 250 instructions,
then it creates over 30 million speculative paths on average per conditional branch.
Often the actual number of paths is smaller because the tree is not balanced, or
because the tree is shallow due to serializing instructions (e.g., system
calls), however the costs are still high, slowing down the fuzzing driver by
orders of magnitude. It could be acceptable for very small fuzzing drivers (e.g., when fuzzing a single function), but not for larger libraries.

The trade-off between the fuzzing speed and the completeness of nested simulation
is a non-trivial one. In particular, it is not clear to what extent added depth of
the simulation improves the detection of speculative vulnerabilities compared to
the loss in input coverage.

To estimate the effectiveness of deeper simulation we compiled our test libraries (see \secref{evaluation}) with \sys{} 
configured for a 6th-order simulation and fuzzed them for 24 hours. 
\tabref{nesting} contains a breakdown of the vulnerabilities we detected by their order.
Clearly, the bulk of the vulnerabilities is detected with only few levels of
nesting, and the higher the order the fewer vulnerabilities we find\footnotemark.

\footnotetext{
The real distribution is even more contrasting.
Here, the 6th-order simulation caused a high overhead and few iterations were executed (\tabref{nesting}).
Therefore, the fuzzer could not generate the inputs to trigger the vulnerabilities with fewer mispredictions.
In fact, in \subsecref{fuzzing-results}, many of these vulnerabilities were discovered by lower-order simulations with more iterations.
}

A plausible explanation of this result is as follows. Most memory accesses are guarded by only one safety check (e.g., a bounds check) which we would need to bypass 
speculatively (first order vulnerabilities). More rarely, the bounds checks would be duplicated across functions or, for example, accompanied by an object type check;
In this case, detecting such a vulnerability would require two mispredictions
(second order). Higher order vulnerabilities usually require the speculative
path to cross several function boundaries.

We can conclude that the speed of fuzzing is a higher priority than the order of simulation.
Most of the vulnerabilities have low orders and we are likely to find more vulnerabilities if we have many iterations of low-order simulation compared to running few iterations of high-order simulation.
In fact, in our later experiments (\subsecref{fuzzing-results}), \sys{} detected more vulnerabilities withing an hour of low-order fuzzing compared to 24 hours with a 6th order simulation.

\myparagraph{Prioritized Simulation}
Based on this observation, we propose the following fuzzing heuristic.
Our \emph{prioritized simulation} tests the low-order paths more rigorously, 
allocating less time to higher-order paths.
%We dedicate different shares of the fuzzing time to different simulation orders, 
%The shares are inversely proportional to the orders.

A simple approach would be to always run the simulation at a baseline order and once every N iterations run a higher-order simulation.
For example, all runs simulate order 1, every 4th run simulates up to order 2, every 16th up to order 3, and so on.

However, since not all runs invoke all the branches, the distribution would be uneven.
Instead, we should calculate the shares per branch.

Suppose we have only two branches---X and Y---in the program under test, and we test the program with six inputs.
X is executed in every run, but Y is invoked only in the runs 1, 2, 3, and 5.
With the prioritized simulation, we simulate only the first-order paths of the branch X in the runs (1, 2, 3, 5, 6) and both the first and the second order paths in the run 4.
As of the branch Y, we simulate the first order in runs (1, 2, 3) and up to the second order in the run 5.

We implemented this strategy in \sys{} and used it in our evaluation.

%\ms{You should explain that the tree is not balanced, so this method does
%not necessarily imply that you always miss exponentially more of deeper nesting
%paths. I don't know if it's true, but it might help convince the reviewers. One nice way to show that would to be to get some statistics (from your
%exhaustive experiments) as for how many paths were really explored.}
%\al{I don't think I can do it.
%Well, I can get the number of traversed paths (though I'd have to re-run the experiment), but, as we discussed last week, we don't know the total number of all possible paths.
%And I haven't come up with a way to estimate it, not yet.}

% placed here for better layout
\begin{figure*}[t]
    \centering
    \includegraphics[scale=0.2]{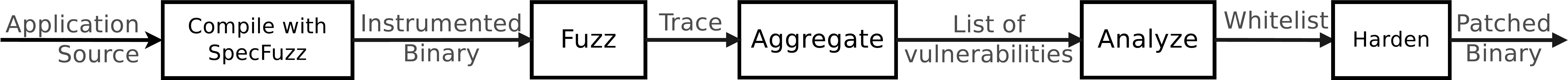}
    \caption{The workflow of testing an application with \sys{}.}
    \label{fig:fuzzing-workflow}
\end{figure*}

\myparagraph{Simulation Coverage}
Because prioritized simulation begins by traversing only one speculative path in every simulation tree and only gradually enters more and more paths, 
it would be important to know which share of all possible speculative paths it managed to cover within a given fuzzing round.
We call this metric a \emph{simulation coverage}. This metric provides an
estimate of the portion of the covered speculated paths out of all possible
paths for all the branches. 

The trade-off different simulation heuristics might explore is a trade-off between
fuzzing coverage and simulation coverage. For example, prioritized simulation gives
preference to the fuzzing coverage. Unfortunately, estimating the precise number
of speculative paths for each branch is a complex problem because the trees are
not balanced. Solving it would require detailed program analysis, which we leave to future work.

\subsection{Other Implementation Details}
\label{subsec:implementation-details}

\myparagraph{External calls and indirect calls}
By the virtue of being implemented as a compiler pass, \sys{} cannot correctly run the simulation beyond the instrumented code.
Therefore, we have to consider all calls to external (non-instrumented) functions as serialization points, even though it is not necessarily a correct behavior (see \secref{discussion}).

Since the complete list of instrumented functions is not known at compile time, \sys{} works in two stages:
It first runs a dummy compilation that collects the function list, and only then does the full instrumentation.
The list can be reused for further compilations if the source does not change.

This approach, however, does not work for indirect calls
%\ms{calls to libraries? because if not libraries, you have it during the compile pass, no?}
%\al{no, indirect call == call through a function pointer. At compile time, we cannot resolve the pointer and thus, we don't know what will be called}
as we do not know the call target at compile time.
Instead, we have to detect the callee type at run time.
To this end, \sys{} inserts a \code{NOP} instruction with a predefined argument into every function entry.
Before indirect calls, it adds a sequence that fetches the first instructions and compares it with the opcode of this \code{NOP}.
If they match, we know that the function is instrumented and it is safe to continue the simulation.

\myparagraph{Callbacks}
There could be a situation where a non-instrumented function calls an instrumented one (e.g., when a function pointer is passed as an argument).
In this case, the instrumented function might return while executing a simulation and the simulation will enter the non-instrumented code, thus corrupting the process state.
To avoid it, \sys{} globally disables simulation before calling external functions and re-enables it afterward.
Accordingly, our current implementation does not support simulation in callbacks (see a potential solution to this problem in \secref{discussion}).

\myparagraph{Long Basic Blocks}
In the end of every basic block (BB), \sys{} checks if the speculation window has expired (i.e., if the instruction counter has reached 250).
This could unnecessarily prolong the simulation when we encounter a long BB, which could be created, for example, by loop unrolling.
To avoid this situation, \sys{} inserts additional checks every 50 instructions in the long BBs.

\myparagraph{Preserving the Process State}
When a function returns while executing a simulation, the value of the stack pointer becomes above its checkpointed value.
Therefore, if we call a function from the \sys{} runtime library or from \asans{}, it would corrupt the checkpointed stack frame.
This could be avoided by logging all changes that these functions do to the memory, but it would have a high performance cost.
Instead, we use a disjoint stack frame for these functions and replace the stack pointer before calling them.

The same applies to the code that \sys{} compiler pass inserts:
We had to ensure that the code that could be executed on a speculative pass never makes any changes to memory besides modifying dedicated variables of the \sys{} runtime.

\myparagraph{Code pointer checks}
Besides causing out-of-bounds accesses, misprediction of conditional branches may also change the program's control flow.
This happens when a corrupted code pointer is dereferenced.
For example, if speculative execution overwrites a return address or the stack pointer, the program can speculatively return into a wrong function or even attempt to execute a data object.
This vulnerability type is especially dangerous as it may allow to launch a ROP-like attack~\cite{Hovav07}.
To detect such corruptions, we insert integrity checks before returning from functions and before executing indirect jumps.
% -*- root: main.tex -*-

\section{Fuzzing with \sys{}}
\label{sec:fuzzing}

%Given the simulation technique described in the previous section (\secref{specfuzz}), we can test applications with conventional dynamic testing methods, such as fuzzing.
%In our experiments,

The workflow is depicted in \figref{fuzzing-workflow}.
\begin{enumerate}
    \item Compile the software under test with Clang and apply the \sys{} pass
    (\secref{specfuzz}), thus producing an instrumented binary that simulates branch mispredictions.
    \item Fuzz the binary. We used HonggFuzz~\cite{HonggFuzz}, an evolutionary coverage-driven fuzzer, and we relied on a combination of custom coverage tracking and Intel Processor Trace~\cite{IntelPT} for measuring coverage.
    \item Aggregate the traces and analyze the detected vulnerabilities to
    produce a \emph{whitelist} of conditional jumps that were deemed
    safe by our analysis.  %executed sufficiently many times during the tests and never triggered an invalid access.
    \item Patch the application with a pass that hardens all but the whitelisted jumps.
\end{enumerate}

We now describe these stages in detail.

\subsection{Coverage and Fuzzing Feedback}
\label{subsec:coverage}

Using existing coverage estimation techniques (e.g., SanitizerCoverage~\cite{sancov}, Intel
PT~\cite{IntelPT}) with \sys{} is incorrect: the values become artificially inflated because \sys{} adds the speculative paths that do not belong to normal program execution.
%Moreover, as mentioned in \subsecref{implementation-details}, the simulation is disabled in callbacks and, thus, we do not want them to be accounted in coverage.

Instead, we implement a custom coverage mechanism that counts executed
conditional branches only outside the speculative paths and when the simulation
is globally enabled (i.e., not in callbacks).
We implement the mechanism through a hashmap that tracks the executed branches as well as the number of unique inputs that triggered every branch.
In addition to coverage, this map is also used for prioritized simulation (\subsecref{nesting}).

We also maintain a hashmap of vulnerabilities as an additional feedback source for evolutionary fuzzing.
This way, every time we detect a new vulnerability, HonggFuzz stores the input that triggered it and adds it to the input corpus.
On top of providing a better feedback to the fuzzer, this feature also allows us to preserve the test cases that trigger specific vulnerabilities.

\subsection{Aggregation of Results}
\label{subsec:aggregation}

As a result of fuzzing, we get a trace of detected speculative out-of-bounds accesses.
Each entry in the trace has a form:
\begin{lstlisting}[style=embedded]
    (Accessed address; Offset; Offending instruction;
    mispredicted branches)
\end{lstlisting}

Here, \emph{offending instruction} is an address of the instruction that tried to access a memory outside the intended object's bounds (\emph{accessed address}), and \emph{mispredicted branches} are the addresses of the mispredicted branches which triggered the access.
\emph{Offset} is the distance to the nearest valid object, if we found one.

To make the trace usable, we aggregate the results per run and per instruction.
That is, for every test run, we collect all the addresses that every unique offending instruction accessed as well as the addresses of the mispredicted branches.

\subsection{Vulnerability Analysis}
\label{subsec:analysis}

After the aggregation, we have a list of out-of-bounds accesses with an approximate range of accessed addresses for each of them.
As we will see in \subsecref{fuzzing-results}, the list may be rather verbose and contain up to thousands of entries.
Yet, we argue that most of them are not realistically exploitable.

In many cases, the violation occurs as a result of accessing an address that remains constant regardless of the program input.
Therefore, the attacker cannot control the accessed address, and cannot leak secrets located in other parts of the application memory.
This could happen, for example, when the application tries to speculatively dereference a field of an uninitialized structure.
In this case, the attacker would be able to leak values from only one address, which is normally not useful unless the desired secret information happens to be located at this address\footnote{
In this work, we do not consider this corner case and leave it to future work.
Its identification would require more complex program analysis (e.g., taint analysis).}.
We call such vulnerabilities \emph{uncontrolled}.

We identify the uncontrolled vulnerabilities by analyzing the aggregated traces.
We estimate the presence of the attacker's control by comparing the accessed addresses in every run (i.e., every new fuzzing input).
If a given offending instruction always accesses the same set of addresses, we assume that the attacker does not have control over it.
Note, however, that the heuristic is valid only after a large enough number of test runs.

After the analysis, we collect a list of safe conditional branches (\emph{whitelist}).
The safety criteria is defined by the user of \sys{}.
In our experiments, the criteria were:
\emph{(i)} the branch was executed at least 100 times;
\emph{(ii)} it never triggered a non-benign vulnerability.
The criteria for defining whether a vulnerability is benign could be controlled too.
In our experiments, they were:
\emph{(i)} the vulnerability was triggered at least 100 times;
\emph{(ii)} the vulnerability is uncontrolled.
In the future, additional criteria could be added to reduce the rate of false positives.

The resulting \emph{whitelist} is a plaint-text file with a list of corresponding code location, which we get based on accompanying DWARF debugging symbols.

\subsection{Patching}
\label{subsec:patch}

Finally, we pass the whitelist created at the analysis stage to a tool that would harden those parts of the application that are not in the list.
We opted for this approach (in contrast to directly patching the detected vulnerabilities) because it ensures that we do not leave the non-tested parts of the application vulnerable.

In our experiments, we used two hardening techniques: adding serializing instructions (\code{LFENCE}s) and adding data dependencies (SLH~\cite{SLH}).

\myparagraph{LFENCE Pass}
The simplest method of patching a BCB vulnerability is to add an \code{LFENCE}---a serializing instruction in Intel x86 architecture that prevents~\cite{intelsys} speculation beyond it.
Adding an \code{LFENCE} after a conditional branch ensures that the speculative out-of-bounds access will not happen.
We used an LLVM pass (shipped as a part of SLH) that instruments all conditional branches with this technique and modified it to accept the whitelist.

\myparagraph{Speculative Load Hardening (SLH)}
An alternative mechanism is to introduce a data dependency between a conditional branch and the memory accesses that follow it.
This mechanism is implemented in another LLVM pass called SLH\@.
We similarly modified the pass to accept the whitelist.

%\myparagraph{Producing minimal patches}
%\al{Under construction}
%\al{I decided not to add it. Too technical and unlikely to be interesting for readers}

\subsection{Investigating Vulnerabilities}
\label{subsec:zoom-in}

Often, it is necessary to go beyond automated analysis and investigate the vulnerabilities manually.
For example, this may be required for penetration testing, for weeding out false positives, or for creating minimal patches where the performance cost of automated instrumentation is not acceptable.

To facilitate the analysis, \sys{} reports all the information gathered during fuzzing.
For vulnerabilities, this information includes:
all accessed invalid addresses and their distance to nearby valid objects (when available);
all sequences of mispredicted branches that triggered the vulnerability;
the order (i.e., the minimal number of mispredictions that can trigger it);
the code location of the fault (based on debug symbols);
whether different inputs triggered accesses to different addresses (controllability);
the execution count.
For branches, the \sys{} reports:
which vulnerabilities this branch can trigger;
the code location of the branch;
its execution count (how many unique inputs covered this branch).

\sys{} also stores the inputs that triggered the vulnerabilities, which could later be used as test cases.

Finally, when the gathered information is not sufficient, \sys{} can instrument a subset of branches instead of the whole application.
This way, we can quickly re-fuzz the locations of interest because such targeted simulation normally runs at close-to-native speed.

% -*- root: main.tex -*-

\section{Evaluation}
\label{sec:evaluation}

% -*- root: ../main.tex -*-
\begin{table}[]
    \center
    \setlength{\tabcolsep}{3.3pt}
    \begin{tabular}{| c | c | c | c | >{\centering}p{1.5cm} |}
        \hline
        MSVC & RH Scanner & Spectector & \sys{} & \bfseries Total \\
        \hline
        7 & 12 & 15 & 15 & \bfseries 15\\
        \hline
    \end{tabular}
    % \vspace{0mm}
    \caption{BCB variants detected by different tools.}
    \label{tab:gadgets}
    % \vspace{-4mm}
\end{table}

In this section, we focus on the following questions:
\begin{itemize}
    \setlength{\parskip}{0pt}
    \setlength{\itemsep}{3pt plus 0pt}
    \item How effective is \sys{} at detecting BCB?
    \item How many vulnerabilities does it find compared to the existing static analysis tools?
    \item How much performance does \sys{} recover over conservative instrumentation of all the branches?%\gr{full-application protection}?
\end{itemize}

\myparagraph{Applications}
We use \sys{} to examine six popular libraries: a cryptographic library (OpenSSL~\cite{OpenSSL} v3.0.0, \code{server} driver),
a compression algorithm (Brotli~\cite{brotli} v1.0.7),
and four parsing libraries, JSON (JSMN~\cite{jsmn} v1.1.0), HTTP~\cite{http} (v2.9.2), libHTP~\cite{libhtp} (v0.5.30), and libYAML~\cite{libyaml} (v0.2.2).
We chose them because they directly process unsanitized input from the network,
potentially giving an attacker the opportunity to control memory accesses within the libraries,
which together with BCB enables random read access to victim's memory by the
attacker.

\myparagraph{Other tools}
To put the results into a context, we compare \sys{} against two existing mitigation and detection tools:
\begin{itemize}
    \setlength{\parskip}{0pt}
    \setlength{\itemsep}{3pt plus 0pt}
    \item RedHat Scanner~\cite{RHScanner}: Spectre V1 Scanner, a static analysis tool from RedHat.
    \item Respectre~\cite{Respectre}: a static analysis tool from GRSecurity.
    Tested only on libHTP as we did not have a direct access to the tool.
\end{itemize}

As a baseline we use LFENCE instrumentation and Speculative Load Hardening (SLH)~\cite{SLH} (shipped with Clang 7.0.1) described in \S\ref{subsec:patch}.

In \subsecref{detection-of-bcb-gadgets}, we additionally tested the \code{/Qspectre} pass of MSVC~\cite{MSVC} (v19.23.28106.4) and a symbolic execution tool Spectector~\cite{Spectector} (commit \code{839bec7}).
Due to low effectiveness, we did not perform further experiments with MSVC\@.
As of Spectector, we report results only for microbenchmarks because larger libraries (Brotli, HTTP, JSMN) exhibited large number of unsupported instructions.

\myparagraph{Testbed}
We use a 4-core (8 hyper-threads) Intel Core i7 3.4\,GHz Skylake CPU, 32\,KB L1
and 256\,KB L2 private caches, 8 MB L3 shared cache, and 32 GB of RAM\@, running Linux kernel 4.16.

\subsection{Detection of BCB Gadgets}
\label{subsec:detection-of-bcb-gadgets}

We tested 15 BCB gadgets by Paul Kocher~\cite{Kocher2018a}.
They were originally designed to illustrate the shortcomings of the BCB
mitigation mechanism in MSVC~\cite{MSVC}. While the suite is not exhaustive,
this is a plausible microbenchmark for the basic detection capabilities.

\tabref{gadgets} shows the results. \sys{} and Spectector expose all speculative out-of-bounds
accesses. MSVC and RedHat~Scanner rely on pattern matching and overlook a few cases.

\subsection{Fuzzing Results}
\label{subsec:fuzzing-results}

To see how effective \sys{} is at detecting vulnerabilities in the wild, we instrumented the libraries with \sys{} configured for prioritized simulation (\subsecref{nesting}) and fuzzed them for varying duration of time: 1, 2, 4, 8, 16, and 32 hours (63 hours in total).
We used one machine and fuzzed on a single thread.
Every next round used the input corpus generated by the previous ones.
The initial input corpus was created by fuzzing the native versions of the libraries for an hour.
Where available, we also added the test inputs shipped with the libraries.

\myparagraph{Fuzzing iterations}
Over the experiment, the average rate of fuzzing was as presented in \tabref{iterations}.
Compared to native, non-instrumented version, \sys{} is definitely much slower.
Yet, the rate is still acceptable:
For example, we managed to test over 400'000 inputs within 63 hours of fuzzing Brotli.

% -*- root: ../main.tex -*-
\begin{table}[]
    \center
    \setlength{\tabcolsep}{3pt}
    \renewcommand{\arraystretch}{1}
    \begin{tabular}{|l|cccccc|}
        \hline
        & JSMN & Brotli & HTTP & libHTP & YAML & SSL\\
        \hline
        Native & 370 & 392 & 463 & 251 & 457 & 84 \\
        \sys{} & 2.8 & 6.6 & 20.4 & 2.4 & 5 & 0.15\\
        \hline
    \end{tabular}\caption{Average number of fuzzing iterations executed by native version and by \sys{} simulation per hour, in thousands.}
    \label{tab:iterations}
\end{table}

\myparagraph{Coverage}
The final coverage of the libraries is shown in \tabref{coverage}.
The presented numbers are branch coverages; that is, which portion of all branches in the libraries was tested during the fuzzing.
We show only the final number (i.e., after 63 hours of fuzzing) because we started with an already extensive input corpus and the coverage was almost not changing across the experiments.
The largest difference was in OpenSSL compiled with \sys{}, where after one hour the coverage was 22.9\% and, in the end, it reached 24\%.

% -*- root: ../main.tex -*-
\begin{table}[]
    \center
    \setlength{\tabcolsep}{3pt}
    \renewcommand{\arraystretch}{1}
    \begin{tabular}{|l|cccccc|}
        \hline
        & JSMN & Brotli & HTTP & libHTP & YAML & SSL\\
        \hline
        Native & 96.6 & 84.1 & 64.1 & 60.6 & 63.9 & 24.0 \\
        \sys{} & 96.6 & 84.1 & 63.5 & 60.6 & 63.3 & 24.0 \\
        \hline
    \end{tabular}\caption{The highest reached coverage of the libraries.
    In percent, out of all branches.}
    \label{tab:coverage}
\end{table}

The difference between the native and the \sys{} versions is caused by our handling of callbacks.
As discussed in \subsecref{implementation-details}, we globally disable the simulation before calling non-instrumented functions.
Hence, some parts of the application are left untested. However, it affects only
performance, not security -- the untested branches remain protected by 
exhaustive instrumentation. 

\myparagraph{Detected Vulnerabilities}
The total numbers of vulnerabilities detected in each experiment is presented in \tabref{detected}.
There is a vast difference between the results, ranging from thousands of violations detected in OpenSSL to only 16 found in the HTTP parser.
The main factor is the code size: OpenSSL has \textasciitilde330000 LoC while HTTP has fewer than 2000 LoC\@.

% -*- root: ../main.tex -*-
\begin{table}[]
    \center
    \setlength{\tabcolsep}{3pt}
    \renewcommand{\arraystretch}{1}
    \begin{tabular}{|l|cccccc|}
        \hline
        Duration & JSMN & Brotli & HTTP & libHTP & YAML & SSL\\
        \hline
        1 hr.  & 20 & 96  & 16 & 322 & 175 & 1940 \\
        2 hr.  & 20 & 101 & 16 & 330 & 202 & 1997 \\
        4 hr.  & 20 & 104 & 16 & 332 & 211 & 2060 \\
        8 hr.  & 20 & 106 & 16 & 334 & 230 & 2104 \\
        16 hr. & 20 & 108 & 16 & 337 & 244 & 2139 \\
        32 hr. & 20 & 108 & 16 & 344 & 251 & 2155 \\
        \hline
    \end{tabular}\caption{Total number of detected vulnerabilities in each experiment.}
    \label{tab:detected}
\end{table}

\myparagraph{Vulnerability types}
For most of the vulnerabilities, however, we did not observe any correlation between the input and the accessed address, which puts them into the category of uncontrolled vulnerabilities (see \subsecref{analysis}).
The results of the analysis are in \tabref{detected-by-type}.
Note that we marked the violations as uncontrolled only if they were triggered by at least 100 different inputs.
Those under the threshold are in the row \emph{unknown}.
\sys{} also detected several cases where the vulnerability corrupted a code pointer (\emph{code}).

% -*- root: ../main.tex -*-
\begin{table}[]
    \center
    \setlength{\tabcolsep}{3pt}
    \renewcommand{\arraystretch}{1}
    \begin{tabular}{|l|cccccc|}
        \hline
        Type & JSMN & Brotli & HTTP & libHTP & YAML & SSL\\
        \hline
        code     &  0  & 2   & 1   & 2    & 3   & 16     \\
        cont.    & 16   &  68  & 9   & 91    & 140   & 589     \\
        uncont.  & 34   &  36  & 6   & 222    & 49   & 1127     \\
        unknown  & 0  & 4  & 0  & 29  & 59 & 423  \\
        \hline
    \end{tabular}\caption{Breakdown of the detected vulnerabilities by type.
    Here, \emph{code} are speculative corruptions of code pointers (e.g., of a return address) and the rest are corruptions of data pointers.
    \emph{Cont.} are controlled vulnerabilities and \emph{uncont.} are uncontrolled.
    \emph{Unknown} are likely uncontrolled vulnerabilities, but they were triggered too few times (less than 100 times).}
    \label{tab:detected-by-type}
\end{table}

%The row \emph{checked} is the number of vulnerabilities that are left after we checked the results for common false positives (i.e., false labeling as controlled).
%The difference between the last two rows is caused by the fact that \asans{} does not provide precise information about the overflows and \sys{} sometimes falsely marks them as \textit{controlled} (see \secref{discussion}).
%We did the checking by using a set of heuristics (implemented as scripts) that identified these cases, which we later verified manually.
%Overall, the manual analysis took 5 hours of work.

%It is still possible that some of the less-typical false positives where not detected.
%Unfortunately, we have not yet developed a method to quickly filter them out.
%With our automated patching approach (\subsecref{patch}), however, such false labeling is acceptable as its only cost is higher overhead.

\myparagraph{Vulnerability orders}
Finally, \tabref{detected-by-order} shows a distribution of the detected vulnerabilities by order.
As we can see, prioritized simulation successfully managed to surface the vulnerabilities up to the 6th order.

% -*- root: ../main.tex -*-
\begin{table}[]
    \center
    \setlength{\tabcolsep}{3pt}
    \renewcommand{\arraystretch}{1}
    \begin{tabular}{|l|cccccc|}
        \hline
        Order & JSMN & Brotli & HTTP & libHTP & YAML & SSL\\
        \hline
        1 & 6 & 79 & 6 & 232 & 97 & 1344 \\
        2 & 7 &  9 & 4 &  66 & 81 &  428 \\
        3 & 5 & 14 & 3 &  33 & 33 &  216 \\
        4 & 2 &  4 & 3 &   5 & 28 &   91 \\
        5 & 0 &  2 & 0 &   6 &  6 &   55 \\
        6 & 0 &  0 & 0 &   2 &  6 &   21 \\
        7 & 0 &  0 & 0 &   0 &  0 &    0 \\
        \hline
    \end{tabular}\caption{Breakdown (by order) of the detected vulnerabilities.}
    \label{tab:detected-by-order}
\end{table}

\subsection{Performance Impact}
\label{subsec:perf}

We used the whitelists produced in the previous experiment to patch the libraries with \code{LFENCE}s and with a modified version of Speculative Load Hardening (see \subsecref{patch}).
Specifically, we used two whitelists for every library: a list based on all out-of-bounds accesses detected by \sys{} and a list that excludes uncontrolled vulnerabilities.

\tabref{removed-branches} shows the shares of the branches that were not
instrumented because of whitelisting (out of the total number of branches in the application).
Naturally, the shares directly correlate with the fuzzing coverage and with the number of detected vulnerabilities.
If the coverage is large, the whitelisting proves to be very effective:
In JSMN, \sys{} reduced the necessary instrumentation by \textasciitilde77\%.

% -*- root: ../main.tex -*-
\begin{table}[]
    \center
    \setlength{\tabcolsep}{1.4pt}
    \renewcommand{\arraystretch}{1}
    \begin{tabular}{|l|cccccc|}
        \hline
        & JSMN & Brotli & HTTP & libHTP & YAML & SSL\\
        \hline
        SLH (all)     & 65\% & 48\% & 44\% & 41\% & 26\% & 15\% \\
        SLH (c,100)   & 69\% & 49\% & 44\% & 50\% & 27\% & 16\% \\
        SLH (c,10)    & 69\% & 49\% & 44\% & 51\% & 37\% & 18\% \\
        LFENCE (all)  & 73\% & 50\% & 56\% & 43\% & 27\% & 16\% \\
        LFENCE (c,100)& 77\% & 51\% & 56\% & 52\% & 28\% & 18\% \\
        LFENCE (c,10) & 77\% & 51\% & 56\% & 53\% & 39\% & 20\% \\
        \hline
    \end{tabular}
    \caption{Shares of branches that avoided instrumentation based on the results of fuzzing.
    \emph{All} means that we patched all detected out-of-bounds accesses, regardless of the type;
    \emph{c,100} means that we did not patch uncontrolled vulnerabilities that were triggered at least 100 times,
    and \emph{c,10}---uncontrolled that were triggered at least 10 times.}
    \label{tab:removed-branches}
\end{table}

Based on these builds, we evaluated the performance impact of the patches.
For the measurements, we used benchmarks included in the libraries, where available;
Otherwise, we used example applications.
As such, we executed: the \code{speed} benchmark in OpenSSL (specifically, RSA, DSA, and ECDSA ciphers);
\code{unbrotli} in Brotli;
\code{bench} in HTTP;
\code{test\_bench} in libHTP;
\code{run-loader} in libYAML;
and a sample parser in JSMN\@.

\begin{figure*}[t]
    \centering
    \includegraphics[scale=0.7]{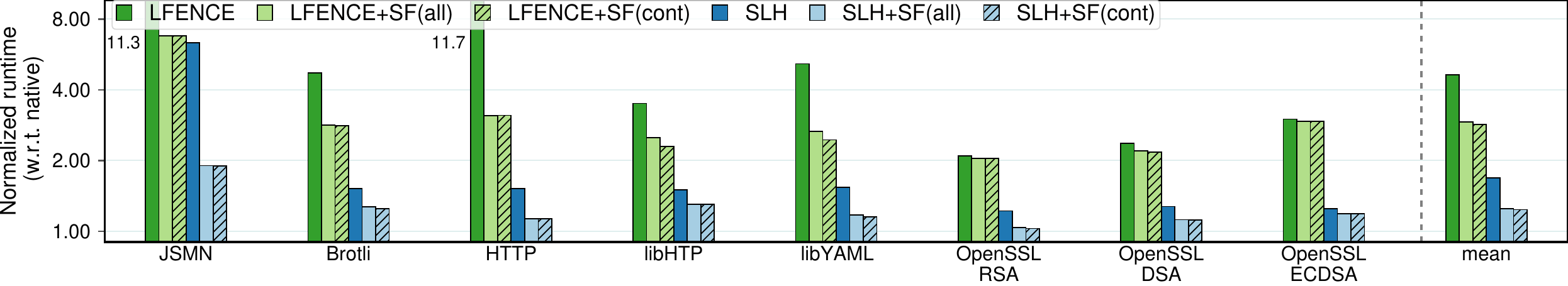}
    \caption{Performance overheads of hardening (Lower is better).
        \emph{+SF(all)} means that we patched all detected out-of-bounds accesses, regardless of the type;
        \emph{+SF(cont)} means that we did not patch uncontrolled vulnerabilities that were triggered at least 100 times.}
    \label{fig:perf}
\end{figure*}

%\begin{figure*}[t]
%    \centering
%    \includegraphics[scale=0.7]{figures/speedup.pdf}
%    \caption{Performance improvement of \sys{}-based patched compared to full hardening (Higher is better). \emph{+SF} are builds with fully-automatic whitelists and \emph{+SF+ch.} are with whitelist based on manually checked vulnerabilities.}
%    \label{fig:speedup}
%\end{figure*}

The results are presented in \figref{perf}.
For clarity, \tabref{perf} shows the same results but interpreted as a speedup of a whitelisted patch compared to full hardening.
As we can see, the overhead is considerably reduced.
The performance cost was, on average, reduced by 23\% for SLH and by 29\% for \code{LFENCE}.

An overall tendency is the higher the coverage of fuzzing, the lower the overhead becomes.
It stems from our benchmarks executing some of the code paths that could not be reached by the fuzzing drivers.

Another parameter is the number and the location of detected vulnerabilities.
In ECDSA, \sys{} detected vulnerabilities on the hot path and, hence, we were not able to remove instrumentation from the places where it caused the highest performance overhead.
\sys{} was also not effective at improving the \code{LFENCE} instrumentation of OpenSSL because it detected speculative bounds violations in the \code{bignum} functions that are located on the hot path.

\begin{table}[]
    \center
    \setlength{\tabcolsep}{3pt}
    \renewcommand{\arraystretch}{1}
    \begin{tabular}{|l|cc|cc|}
        \hline
        & \multicolumn{2}{c|}{SLH} & \multicolumn{2}{c|}{LFENCE} \\
         & +SF(all) & +SF(cont) & +SF(all) & +SF(cont) \\
        \hline
        JSMN    &  233\% & 234\% & 131\% & 132\% \\
        Brotli  &   20\% &  22\% &  66\% &  67\% \\
        HTTP    &   34\% &  34\% & 243\% & 242\% \\
        libHTP  &   15\% &  15\% &  40\% &  52\% \\
        YAML    &   30\% &  33\% &  93\% & 110\% \\
        RSA     &   17\% &  19\% &   2\% &   2\% \\
        DSA     &   13\% &  14\% &   8\% &   9\% \\
        ECDSA   &    5\% &   5\% &   2\% &   2\% \\
        \hline
    \end{tabular}\caption{Performance improvement of \sys{}-based patches compared to full hardening.
        \emph{+SF(all)} means that we patched all detected out-of-bounds accesses, regardless of the type;
        \emph{+SF(cont)} means that we did not patch uncontrolled vulnerabilities that were triggered at least 100 times.}
    \label{tab:perf}
\end{table}

A major reasons for relatively high overheads is an issue with debug symbols that we encountered in LLVM\@.
Sometimes, the debug symbols of the same code location would mismatch between compilations with different flags or would be completely absent for some instructions.
Accordingly, some of the whitelisted locations would still be hardened.
Note that this bug only impacts the performance, not the security guarantees.
Nevertheless, when the issue is resolved, the overheads are likely to get lower.

One interesting example is JSMN, which experienced 5x slowdown with SLH and 11x with the \code{LFENCE} instrumentation.
It is caused by an extremely high density of branches in the application (approximately one branch executed every cycle) and, thus, high reliance on branch prediction to efficiently utilize instruction parallelism.
Complete hardening effectively disables this optimization and makes the execution much more sequential.
At the same time, \sys{} found very few vulnerabilities in JSMN and had high coverage (96\%).
Hence, the patches improved the performance by 230\% (\code{LFENCE}) and 130\% (SLH)

\subsection{Comparison with Other Tools}
\label{subsec:comparison}

\myparagraph{Spectre Scanner}
For comparison, we also tested the libraries with RedHat Scanner (\tabref{rhs}).
Although it detected fewer vulnerabilities than \sys{}, it found many vulnerabilities that \sys{} did not (second row).
The reason behind it is almost all of them were located in the parts of code not covered during fuzzing.
There were only two exceptions (row three), but both turned out to be false positives.
(Because of the overwhelming amount of data, we did not investigate which share of the second row were false positives).

%Note that when the whitelist-based patching is applied (\subsecref{patch}), these undetected vulnerabilities do not compromise the security guarantees of \sys{}.

\myparagraph{Respectre}
Thanks to a cooperation with GRSecurity, we were able to also compare our results to a commercial static analysis tool Respectre~\cite{Respectre}.
As a test case we selected libHTP\@.
In total, Respectre detected 167 vulnerabilities, out of which \sys{} found 79.
Similarly to the previous experiment, the other 88 are located in the parts of libHTP not covered by fuzzing.

\sys{} was able to detect more vulnerabilities due to its more generic nature:
For example, it can detect vulnerabilities that span multiple functions.
On the other hand, Respectre is not confined by coverage and it can detect vulnerabilities in the parts of the application that cannot be reached by fuzzing.

\reportonly{
Note, however, that these numbers are based on the raw outputs of both tools.
It is possible that they include false positives.
We attempted to manually analyze them, but the task appeared to be too complex.
Speculative vulnerabilities in real software constitute of complex interactions of correct and wrong predictions, and analyzing them by just looking at the code has proven to be too error prone.
Therefore, we report only raw numbers and leave the analysis of false positives to future work (e.g., to program analysis tools, as discussed in \secref{discussion}).
}

% -*- root: ../main.tex -*-
\begin{table}[]
    \center
    \setlength{\tabcolsep}{2.6pt}
    \renewcommand{\arraystretch}{1}
    \begin{tabular}{|l|cccccc|}
        \hline
        Order & JSMN & Brotli & HTTP & libHTP & YAML & SSL \\
        \hline
        Both & 1 & 6 & 1 & 78 & 3 & 992 \\
        RHS & 0 & 4 & 3 & 36 & 3 & 601 \\
        RHS/covered & 0 & (1) & 0 & 0 & 0 & (1) \\
        \hline
    \end{tabular}
    % \vspace{0mm}
    \caption{Vulnerabilities detected by \sys{} and RH Scanner.
    The first row are the vulnerabilities detected by both tools;
    the second---only by RH Scanner;
    the third row are the vulnerabilities detected only by RH Scanner and located on the paths covered during our fuzzing experiments.}
    \label{tab:rhs}
\end{table}

%\al{I gave up on trying to do a detailed comparison. When I try to analyse specific vulnerabilities that one or another tool found, the question always comes up: Is it really a vulnerability or a false positive? Given the amount of data, it's virtually impossible to answer this question with any significant confidence. Therefore, I'll keep it short. At least, for this submission.}

%\input{tables/respectre}

% \myparagraph{Spectector}
% \al{The experiments are running}

\subsection{Case Studies}
\label{subsec:casestudy}

In this section, we present a detailed overview of three potential vulnerabilities found by \sys{}.
Note that we did not test them in practice.

\myparagraph{Speculative Overflow in libHTP base64 decoder} One of the utility functions that libHTP provides is base64 decoder, which is used to receive user data or parameters that may be sent in text format.
This functionality is implemented in function \code{htp\_base64\_decode}, which calls function \code{base64\_decode\_single} in a loop.
\code{base64\_decode\_single} decodes a Base64 encoded symbol by looking it up in a table of precomputed values (array \code{decoding}, lines 2--3).
Before fetching the decoded symbol, the function checks the value for over- and underflows.
The attacker can bypass the check by training the branch predictor and, thus, trigger a speculative overread at line 7.

Two properties make this vulnerability realistically exploitable.
First, the attacker has control over the accessed address because the array index (\code{value\_in}) is a part of the HTTP request.
Second, the fetched value is further used for defining the control flow of the program (see the comparison at line 16), which allows the attacker to infer a part of the value (specifically, its sign) by observing the cache state.

The attacker could execute the attack as follows.
She begins by sending a probing message to find out which cache line the first element of the array \code{decoding} uses.
Then, she sends a valid message to train the branch predictor on predicting the bounds check (line 5) as true.
Finally, she resets the cache state (e.g., flushes the cache) and sends a message that contains a symbol that triggers an overread, followed by a symbol that triggers a read from the first array element.
If the read value is negative, the loop will do one more iteration, execute the second read, and the attacker will see a change in the state of the corresponding cache line.
Otherwise, the loop will be terminated and the state will not change.

\myparagraph{Speculative Overflow in OpenSSL ASN1 decoding}
Another vulnerability is in OpenSSL ASN1 decoder.
It is used to decode, for example, certificates that clients send to the server.

The attacker sends malicious ASN1 data to the victim.
The victim uses \code{asn\_$*$\_d2i} family of functions to parse the message.
One of the functions is \code{asn1\_item\_embed\_d2i}, which, among others, decodes components of type \code{MSTRING}, verifying its tag in the process.
The tag of the message is extracted through a call to \code{asn1\_check\_tlen} function, which delegates this calculation to \code{ASN1\_get\_object}.
\code{asn1\_check\_tlen} verifies if the received tag matches the expected one (lines 22 and 23), however a misspeculation on any of these lines can nullify this check.
Later, \code{asn1\_item\_embed\_d2i} calls \code{ASN1\_tag2bit} on the decoded tag value.
If misspeculation happens in this function as well (line 4), the array \code{tag2bit} will be indexed with a potentially unbounded 4-byte integer.
Later, this value is used to derive the control flow of the application (line 14), which may be used to leak user information.

\myparagraph{Jump address corruption in OpenSSL ASN1}
\sys{} detected a vulnerability that may speculatively change the control flow of the program in \code{asn1\_ex\_i2c}.
This function includes a switch statement with a tight range of values.
Such switches are often compiled as jump tables (if this optimization is not disabled explicitly).

A misprediction in the switch statement may cause an out-of-bounds read from the jump table.
Accordingly, a later indirect jump would dereference a corrupted code pointer and the program will jump into a wrong location.
In our experiments, we saw it jumping into the functions that were nearby in the binary (e.g., into \code{asn1\_primitive\_free}), but, with careful manipulation of the object and data layouts, this may be extended to a speculative ROP attack.

\begin{figure}[t]
   \begin{lstlisting}[frame=tb]
int base64_decode_single(signed char value_in) {
 static signed char decoding[] =
   {62, -1, ...}; // 80 elements
 value_in -= 43;
 if ((value_in < 0) || (value_in > decoding_size - 1))
   return -1;
 return decoding[(int) value_in];
}
...
int htp_base64_decode(const void *code_in, ...) {
 signed char fragment;
 ...
 do {
   ...
   fragment = base64_decode_single(*code_in++);
 } while (fragment < 0);
...}\end{lstlisting}
   \caption{A BCB vulnerability in a Base64 decoding function.}
   \label{fig:htp-base64}
\end{figure}

\begin{figure}[t]
   \begin{lstlisting}[frame=tb]
const unsigned long tag2bit[32] = {...};
unsigned long ASN1_tag2bit(int tag) {
    // misspeculation required
    if ((tag < 0) || (tag > 30)) return 0;
    return tag2bit[tag];
}
int asn1_item_embed_d2i(ASN1_VALUE **pval, ...) {
    int otag;
    ...
    switch (it->itype) {
    case ASN1_ITYPE_MSTRING:
        ret = asn1_check_tlen(..., &otag, ...);
         ...
        if (!(ASN1_tag2bit(otag) & it->utype)) {...}
    }
}
int asn1_check_tlen(..., int *otag, int expclass) {
  ...
  // decodes the ptag from message
  i = ASN1_get_object(..., &ptag);
  ...
  if (exptag >= 0) {
    if ((exptag != ptag) || (expclass != pclass)) {
      // misspeculation required
  ...
\end{lstlisting}
   \caption{A BCB vulnerability in a ASN1 decoding function.}
   \label{fig:openssl-asn1}
\end{figure}

%%% Local Variables:
%%% mode: latex
%%% TeX-master: "ms"
%%% End:

% -*- root: main.tex -*-

\section{Other Spectre Attacks}
\label{sec:apply-simul-spectre}

Bounds Check Bypass is not the only type of speculative vulnerabilities that could be detected by speculative exposure.
Below we give an overview of instrumentation that can be used for other Spectre-type attacks.

\myparagraphnodot{Branch Target Injection}~\cite{Kocher18} is a Spectre variant targeting speculation at indirect jumps.
When an indirect jump instruction is executed, the CPU speculates the jump target using the branch predictor without waiting for the actual target address computation to finish.
The attacker can exploit this behavior by training the branch predictor to execute a jump to a code snippet that would leak program data via a side channel.

\sys{} could be modified to simulate BTI by maintaining a software history buffer for every indirect branch in the application.
Then, at an indirect branch, \sys{} would \emph{(i)} record the current branch target into the history buffer and \emph{(ii)} run a simulation for every previously recorded target.
This approach works, however, only under the assumption that attacker can train the branch predictor only by providing data to the application and cannot inject arbitrary targets into the branch predictor's history buffer from another application on the same core.

\myparagraphnodot{Return Address Misprediction}~\cite{Maisuradze2018, Koruyeh2018} attack is a variant of Branch Target Injection.
The CPU maintains a small number of most recently used return addresses in a dedicated cache, pushing the return address into this cache on each call instruction and popping it from the cache on each return instruction.
When this cache becomes empty, the CPU will speculate the return address using the indirect Branch Target Buffer.
To simulate this vulnerability, \sys{} could instrument call and return instructions to, correspondingly, increment and decrement a counter, jumping to an address from history buffer on return addresses with negative or zero counter value.
This simulation should be combined with the previous one as the return address prediction could fall back to indirect branch target prediction.
\al{A reviewer complained that this paragraph is hard to understand (I agree with it). Let's rewrite it}

\myparagraph{Speculative Store Bypass}~\cite{Google2018} is a microarchitectural vulnerability caused by CPU ignoring the potential dependencies between load and store instructions during speculation.
When a store operation is delayed, a subsequent load from the same address may speculatively reuse the old value from the cache.
To simulate this attack, \sys{} could be extended to start a simulation before every write to memory.
Then, \sys{} would skip the store during the simulation, but execute it after the rollback.

\section{Limitations}
\label{sec:discussion}

In this section, we discuss the conceptual problems we have discovered while developing \sys{} as well as potential solutions to them.

\myparagraph{Reducing the Complexity of Nested Simulation}
As we discussed in \subsecref{nesting}, complete nested simulation is too expensive and limiting the order of simulation may lead to false negatives.
One way we could resolve this problem is by statically analyzing the program before fuzzing it, such that the typical vulnerable patterns as well as typical false positives would be purged from the simulation, thus reducing its cost.
%This way, we get the best from both domains: the speed of static analysis tools and the universality of \sys{}.

\myparagraph{False Negatives}
\sys{} will not find a vulnerability if the fuzzer does not generate an input that would trigger it.
Unfortunately, it is an inherent problem of fuzzing.

\myparagraph{Fuzzing Driver}
Another inherent issue of all fuzzing techniques is their coverage.
As we saw in \secref{evaluation}, it highly depends on the fuzzing driver and a bad driver may severely limit the reach of testing.
Since we use whitelist-based patching, low coverage may cause high performance overhead in patched applications.
It could be improved by applying tools that generate drivers automatically, such as FUDGE~\cite{fudge}.

\myparagraph{Mislabeling}
During the evaluation, we discovered that our vulnerability analysis technique (see \subsecref{fuzzing}) sometimes gives a false result and mistakenly labels an uncontrolled vulnerability as a controlled one.
It happens because \asan{} reports only the accessed address and not the distance between the address and the referent object (i.e., offset).
Therefore, if the object size differs among the test runs, the accessed address will also be different, even if the offset is the same.

For example, one common case of mislabeling is off-by-one accesses.
If an array is read in a loop, our simulation will force the loop to take a few additional iterations and read a few elements beyond the array's bounds.
If the array size differs from one test run to another, the analysis would mark this vulnerability as controllable.

%So far, we handled small overflows of this type by checking if any of the close neighboring addresses near the invalid access contain a valid object.
%Yet, for larger overflows we had to resort to manually filtering the false positives.

To avoid this issue, we could use a more complete memory safety technique (e.g., Intel MPX~\cite{intelsys}) that maintains metadata about referent objects.
%That would allow us to filter by changes in the distance to the object bound instead of changes in the accessed address.
Unfortunately, none of such techniques is supported by LLVM out-of-the-box.
To resolve this issue, we would have to implement MPX support or migrate \sys{} to another compiler.

An even better solution would be to use a program analysis technique (e.g., taint analysis or symbolic execution) to verify the attacker's control.
We leave it to future work.

\myparagraph{Legacy Code and Callbacks}
Because we implemented \sys{} as a compiler pass, it cannot run the simulation in non-instrumented parts of the application (e.g., in system libraries) as well as in the calls from these parts (callbacks).
To overcome this problem, we could have implemented \sys{} as a binary instrumentation tool (e.g., with PIN~\cite{luk2005pin}).
Yet, techniques of this type are normally heavy-weight and it would considerably increase the required fuzzing time.
%\todo{add citation}
% -*- root: main.tex -*-

\section{Related Work}
\label{sec:related}

%BCB is a combination of three vulnerabilities: a branch misprediction, a bounds violation, and a side-channel leakage.
%Accordingly, three types of defenses exist.

%\myparagraph{Preventing Unsafe Speculation}
%\label{subsec:preventing-unsafe-speculation}
The most conservative solution to Spectre-type attacks is to disable prediction entirely~\cite{SUSE} (although not all processors support it) or on a targeted basis, with serializing instructions (e.g., \code{LFENCE} on Intel CPUs or \code{DSBSY} on ARM\@).
Speculation can also be delayed by adding a data dependency, as implemented in SLH~\cite{SLH} and YSNB~\cite{YSNB}).
As we saw in \secref{evaluation}, it causes a considerable slowdown.

Static analysis is often used to detect the Spectre-type vulnerabilities and avoid the high performance cost of full hardening.
Tools like Spectre 1 Scanner~\cite{RHScanner}, MSVC Spectre 1 pass~\cite{MSVC}, and Respectre~\cite{Respectre} analyze the binary and search for Spectre gadgets.
Although mature tools like Respectre can detect many vulnerabilities (see~\secref{evaluation}), the reliance on predefined patterns may leave an unexpected variant to stay unnoticed.

Alternatively, oo7~\cite{Wang2018} relies on static taint analysis to detect the memory accesses that are dependent on the program input.
(This is the same criteria that we used to identify uncontrolled vulnerabilities.)
%Specifically, oo7 searches for the following pattern:
%A conditional branch with a condition dependent on the input (i.e., tainted) is followed by a load dependent on the condition, followed by memory access dependent on the load.
This approach is more universal than the pattern-matching techniques, but it is affected by the inherent problems of static taint analysis.
Namely, limited analysis depth may cause false positives and overtainting causes false negatives.

Tools like Spectector~\cite{Spectector}, Pitchfork~\cite{pitchfork}, and SpecuSym~\cite{SpecuSym} apply symbolic execution to detect Spectre-type vulnerabilities.
Although they often provide stronger security guarantees compared to fuzzing, an inherent problem of symbolic execution is combinatorial explosion, which is further exacerbated by nested speculation.

A long-term solution to the problem lays in modifications to the hardware.
InvisiSpec~\cite{Yan2018} and SafeSpec~\cite{Khasawneh2018} propose separate hardware modules dedicated to speculation.
CleanupSpec~\cite{CleanupSpec} cleanses the cache traces when a misprediction is detected.
NDA~\cite{NDA} restricts speculation to only ``safe'' paths.
Context-Sensitive Fencing~\cite{Taram2019} inserts serialization barriers at decoding stage upon detecting a potentially dangerous instruction pattern.
ConTExT~\cite{Schwarz2019} proposes an extension to the memory management mechanism that isolates safety-critical data.
These techniques, however, do not protect the existing processors vulnerable to Spectre-type attacks.

%TODO: https://repo.or.cz/w/smatch.git

%TODO: Sakalis et al.~\cite{Sakalis2019} - I have to read this one carefully. Not exactly sure what they do.

%TODO: SpectreGuard, ScatterCache, Identifying Cache-Based Side Channels through Secret-Augmented Abstract Interpretation, the code that never ran

%\myparagraph{Preventing Memory Safety Violations}
%\label{subsec:preventing-memory-safety-violations}
Classical memory safety techniques (e.g., Intel MPX~\cite{intelsys}, SoftBound~\cite{softbound09}) do not protect from BCB, but can be retrofitted to disable speculative accesses.
A variant of it---index masking---is now used in JavaScript engines~\cite{Wagner2018} where, before accessing an array element, the index is masked with the array size.
As it is an arithmetic operation, it does not create a control hazard and is not predicted by the CPU\@.
However, this defense is vulnerable to the attacks where the data type is mispredicted and a wrong mask is used~\cite{Alephsecurity2019}.

%\myparagraph{Preventing Side Channels}
%\label{subsec:preventing-side-channels}
Another approach is to eliminate the possibility of leaking speculative results through a side channel (SC).
There is an extensive body of research in this direction, ranging from cache isolation~\cite{Sprabery2018,DAWG}, to attack detection~\cite{Gruss2017},
enforcing non-interrupted execution~\cite{Varadarajan2014b, Varys}, and cache coloring~\cite{shi2011limiting}.
Yet, they protect only against specific SC and speculative attacks may use various channels~\cite{schwarz2018netspectre}.
A relatively complete isolation can be achieved with a specialized microkernel~\cite{ge2019time}, but it requires a complete system redesign.

In practice, browsers mitigate SCs by reducing the resolution of timers~\cite{Wagner2018}, disabling shared memory or using site isolation~\cite{SiteIsolation}.
These techniques prevent only cross-site attacks, and are not effective at the presence of a local attacker.

% -*- root: main.tex -*-
\vspace{-0.5cm}

\section{Conclusion}
\label{sec:conclusion}

We presented a technique to make speculative execution vulnerabilities visible by simulating them in software.
We demonstrated the technique by implementing a Bounds Check Bypass detection tool called \sys{}.
During the evaluation, the tool has proven to be more effective at finding vulnerabilities than the available static analysis tools and the patches produced based on the fuzzing results had better performance than conservative hardening techniques.

Yet, this work is only a first attempt at applying dynamic testing techniques to detection of speculative execution vulnerabilities.
We hope that it will show the promise of this research direction and will help pave the way for future, even more efficient vulnerability detection tools.

%\section*{Availability}
%\label{sec:github}
\myparagraph{Availability}
Source code of \sys{} is publicly available under \url{https://github.com/tudinfse/SpecFuzz}.
%\vspace{-0.2cm}

%\section*{Acknowledgments}
%\label{sec:acks}
%\vspace{-0.2cm}
\myparagraph{Acknowledgments}
This work was funded by
the Federal Ministry of Education and Research of the Federal Republic of Germany (03ZZ0517A, FastCloud);
the EU H2020 Programme under the LEGaTO Project (legato-project.eu), grant agreement No. 780681;
and with support from the Technion Hiroshi Fujiwara Cybersecurity.

\balance

\bibliographystyle{plain}
\bibliography{ms}

%\appendix
%\input{91_appendix_trace_example}
%\input{92_terminology}
%\input{93_major_revision}

% \theendnotes

\end{document}